\documentclass[12pt,onecolumn]{IEEEtran}
\ifCLASSINFOpdf
   \usepackage[pdftex]{graphicx}
\else
\fi

\usepackage{cite}
\usepackage{amsmath}
\usepackage{array}
\usepackage{lineno,hyperref}

\usepackage[usenames, dvipsnames]{color}
\usepackage[table]{xcolor}
\usepackage[linesnumbered,ruled,vlined]{algorithm2e}

\usepackage{algorithmicx}
\usepackage{algpseudocode}
\usepackage{amssymb}
\usepackage{xr}
\usepackage{multirow}
\usepackage{hhline}
\usepackage{array}
\usepackage{fancyvrb} 
\usepackage{multicol}

\usepackage[textsize=scriptsize]{todonotes}
\usepackage[normalem]{ulem}

\MakeRobust{\Call}

\usepackage{amsmath}               
{

}
\usepackage{fixltx2e}
\usepackage[normalem]{ulem}
\graphicspath{{pics/}}
\usepackage[justification=centering]{caption} 

\usepackage{threeparttable}
\usepackage{multirow}
\usepackage{comment}

\usepackage{balance}
\usepackage{flushend}

\begin{document}
\title{EXCLUVIS: A MATLAB GUI Software for Comparative Study of Clustering and Visualization of Gene Expression Data} 

\author{Sudip Poddar, Anirban Mukhopadhyay
\thanks{Sudip Poddar is with the Institute for Integrated Circuits, Johannes Kepler University Linz, Austria (E-mail: sudippoddar2006@gmail.com). Anirban Mukhopadhyay is with the Department of Computer Science \& Engineering, University of Kalyani, Kalyani-741235, India (E-mail: anirban@klyuniv.ac.in). The main content of this article is published from the Master’s thesis of Sudip Poddar submitted to University of Kalyani, West Bengal, India in June 2012. Copyright©2020, Retained by the authors. All rights reserved.}
}

\maketitle
\IEEEpeerreviewmaketitle

\begin{abstract}
Clustering is a popular data mining technique that aims to partition an input space into multiple homogeneous regions. There exist several clustering algorithms in the literature. The performance of a clustering algorithm depends on its input parameters which can substantially affect the behavior of the algorithm. Cluster validity indices determine  the partitioning that best fits the underlying data. In bioinformatics, microarray gene expression technology has made it possible to measure the gene expression levels of thousands of genes simultaneously. Many genomic studies, which aim to analyze the functions of some genes, highly rely on some clustering technique for grouping similarly expressed genes in one cluster or partitioning tissue samples based on similar expression values of genes. In this work, an application package called {\bf EXCLUVIS} (gene EXpression data CLUstering and VISualization) has been developed using {MATLAB} Graphical User Interface (GUI) environment for analyzing the performances of different clustering algorithms on gene expression datasets. In this application package, the user needs to select a number of parameters such as internal validity indices, external validity indices and number of clusters from the active windows for evaluating the performance of the clustering algorithms. {\bf EXCLUVIS} compares the performances of $K$-means, fuzzy $C$-means, hierarchical clustering and multiobjective evolutionary clustering algorithms. Heatmap and cluster profile plots are used for visualizing the results. {\bf EXCLUVIS} allows the users to easily find the goodness of clustering solutions as  well as provides visual representations of the clustering outcomes.
\end{abstract}

\begin{IEEEkeywords}
Bioinformatics, microarray, gene expression, clustering, validity indices, graphical user interface, package, algorithm, MATLAB.
\end{IEEEkeywords}
\IEEEpeerreviewmaketitle

\section{Introduction}

Clustering is an important unsupervised data mining task that partitions the input space into different homogeneous clusters such that the objects within the same cluster are as similar as possible, while the objects belonging to different clusters are as dissimilar as possible~\cite{jain99}. The similarities among the objects are measured in terms of some distance metric. There exist a number of clustering techniques in literature such as partitional clustering, hierarchical clustering, density-based clustering, and
evolutionary algorithm-based clustering~\cite{mukhopadhyay2015}. 

Microarray gene expression datasets are useful for studying the expression levels of thousands of genes simultaneously \cite{quack01,shannon03}.
Clustering gene expression data helps in grouping the genes based on their expression patterns or grouping the samples based on the gene expression values, 
which further facilitates prediction of gene functions and genetic markers. Since there exist a number of clustering algorithms in literature,
therefore a software for comparing these algorithms for a particular expression dataset will be very helpful for the biologists and bioinformaticians. 

In view of this, we have developed a {MATLAB} \cite{MATLAB} GUI package called {\bf EXCLUVIS} (gene EXpression data CLUstering and VISualization) for comparative study of clustering and visualization of gene expression data. The package, developed in {MATLAB 2009b}, presents a very user-friendly graphical interface for comparison of different clustering algorithms visually and numerically. In this initial version, we have implemented some popular clustering algorithms like $K$-means, fuzzy $C$-means, hierarchical clustering algorithms and a multiobjective clustering algorithm \cite{uas09}. However, one can incorporate other clustering algorithms in future as the package is open-source. The performances of the clustering algorithms can be compared in terms of some cluster validity indices and also by visualization of the clustering results.   

In the subsequent sections we describe data preprocessing techniques, visualization tools, clustering algorithms used in this software, cluster validity indices incorporated and system requirement followed by demonstration of the software. Finally we discuss the availability of the software package and conclude the article.

\section{Data preprocessing}
Data preprocessing is an important step in data analytics. The primary step in data analytics is data collection to form a data matrix. The collected data may be dirty i.e. may contain noise, garbage values, null values, incomplete, inconsistent, duplicated. So, to make the collected data usable, there is a need to preprocess the data before making it usable. The preprocessing task has two components: data cleaning and data transformation. Data cleaning is the process of detecting and correcting the irrelevant data in the data matrix. The data transformation is the process (i) to transform all the categorical values to numeric values (integer or real) such that all the entries must be in numeric form and (ii) to normalize all the entries in the data matrix in a common range (e.g. [0,1] or [-1,1]). After data transformation, the analysis of features is the next task in data analytics. In {\bf EXCLUVIS}, we have first calculated the variances of the genes and then sorted then in decreasing order of the same.

\section{Visualization}

Groups of functionally related genes in microarray data can be identified by applying the available clustering algorithms in data mining literature.  But it is very difficult to find out the most appropriate algorithm to apply 
due to the lack of a gold-standard verification of any clustering algorithm. Interestingly, this analytical process can also be performed easily using data visualization tools such as heatmap, profile plot etc. For this reason, 
these two features are included in this application package for visualizing the expression level of various genes.

\subsection{Heatmap}
Heatmap is considered as a widely used popular data visualization technique, which plots the genomic data in a two dimensional grid.
It is generally used to visualize gene expression data in which the rows correspond to the genes and the columns correspond to the features. Here, 
the magnitude of each matrix entry is represented using a color scale. Hence, heatmap provides a generalized view of data in colored representation.

\subsection{Cluster profile plot}

In gene expression analysis, the expression profile of a gene is studied in different experimental conditions. In multiple phenotype conditions, 
perturbation in expression pattern is detected by visualizing the expression profiles. Another complementary approach to visualize the dynamics 
of altered expression patterns is to measure the gene expression at different time interval, under the same phenotype condition. Finding genes with
similar expression pattern is one of the main interests for the biologists, as these genes provide a mean to understand the co-regulation pattern in a
gene network. Several methods are adopted from machine learning and statistics to find co-expressed/co-regulated genes. Cluster profile plot is 
used to visualize those groups of co-regulated genes and it is one of the popular visualization tools for the biologists.

\section{Clustering algorithms}

In this application package several clustering algorithms are integrated for identifying groups of functionally related genes in microarray data. The results of clustering solution are validated using validity indices. Also this package enable the users to visually compare clustering solutions using heatmap and profile plot. Clustering algorithms that have been implemented in {\bf EXCLUVIS} are described in the following sections.

\subsection[K-means]{$K$-Means}

In statistics and data mining, $K$-means clustering is widely used clustering technique developed by MacQeen in 1967 \cite{kmeans}.
It is one of the simplest and effective techniques that aims to partition $n$ observations into $K$ clusters in $d$-dimensional space.
The partitioning is performed by assigning each observation to the nearest mean. It minimizes a squared error as objective function defined as follows:
\begin{equation}
J = \sum_{j = 1}^{K}\sum_{X_i \in C_j}^{n}\parallel{X_{i} - c_{j}}\parallel^ 2,
\end{equation}
where $\parallel{X_{i} - c_{j}}\parallel $ is a chosen distance measure between a data point $X_{i}$ and cluster center $c_j$. $K$-means minimizes the global 
cluster variance $J$ to maximize the compactness of the clusters. It may happen that the values returned by $K$-means is not optimal and for fixed $K$ and $d$, 
this can be solved in $O(n^{dK+1}\log n)$ time, where $n$ is the number of entities need to be clustered.

\subsection[Fuzzy C-means]{Fuzzy $C$-Means}

In fuzzy $C$-means (FCM) clustering~\cite{bez81cmeans,jc73}, each observation belongs to a cluster with a certain degree of membership value. 
This method (developed by Dunn in 1973 and improved by Bezdek in 1981) is widely used in statistics and pattern recognition. It is based on minimization of the following objective function:
\begin{equation}
J_m = \sum_{i = 1}^{K}\sum_{j = 1}^{n} u_{ij}^{m}\parallel{X_{j} - c_{i}}\parallel^2,
\end{equation}
where $m$ is any real number greater than 1, $X_j$ is the $j$th of $d$-dimensional measured data, $ u_{ij}$ is the degree of membership of $X_j$ in the cluster $i$, $c_i$ is the $d$-dimensional center of cluster $i$.
FCM generally produces better clustering results than $K$-means for overlapping clusters and noisy data. However, both FCM and $K$-means are sensitive to outliers.

\subsection{Hierarchical clustering}

In Hierarchical clustering \cite{sc67,rd78}, a sequence of clusters is  generated in a hierarchy. Each level of hierarchy provides a particular
clustering of the data. Hierarchical clustering may be either agglomerative or divisive. In agglomerative clustering, at first each data point is 
regarded as a singleton cluster. In each iteration, two nearest clusters are merged into a single cluster. The merging is performed until a single cluster remains.
On the contrary, in divisive case it starts with a single cluster containing all the data points. At each step, clusters are successively split into smaller clusters according to some dissimilarity measure.

The main shortcoming of hierarchical clustering is that the interpretation of the hierarchy is complex and often confusing. The deterministic nature 
of the method prevents reevaluation of the clusters after grouping the nodes. 
Also the time complexity is at least $O(n^2)$, where $n$ is the total number of objects and they can never undo what is already done.
In {\bf EXCLUVIS}, some agglomerative hierarchical clustering algorithms are implemented.

\subsection{Multiobjective clustering with support vector machine (MocSvm)}

A multiobjective evolutionary algorithm-based clustering algorithm \cite{au09} is also included into {\bf EXCLUVIS}. 
In this approach, two cluster validity indices, namely $J_m$ index \cite{bez81cmeans} and Xie-Beni index \cite{xlg91} are optimized
simultaneously to yield robust clustering solutions. The algorithm is developed based on non-dominated sorting genetic
algorithm II (NSGA-II) \cite{deb02,mukhopadhyay2015} and generates a near-Pareto-optimal set of clustering solutions.
These solutions are then integrated based on a fuzzy majority voting with support vector machine (SVM) classifier to obtain a single final solution.

\section{Cluster validity indices}

The main objective of clustering is to find similar groups of objects present in a dataset. Each clustering algorithm searches for clusters 
in which members are close to each other showing high degree of similarity.  The main difficulty of clustering algorithms is to find the optimal number of clusters that best suits the dataset. 
Note that while visual verification of the validity of clustering results in $2D$ or $3D$ data set may be possible, for multidimensional data it is very difficult to validate the clustering results visually. Moreover the clustering results may produce non-optimal number of clusters for improper value of parameters. The problem of finding the optimal number clusters and visualization of clustering results has been subjected to several research efforts.
In general, there are two approaches to investigate cluster validity.

\begin{itemize}
	\item External Indices: Used to measure the extent to which cluster labels match with externally supplied class labels, e.g., Minkowski index.
	\item Internal Indices:  Used to measure the goodness of a clustering structure based on the intrinsic information of the data alone, e.g., Sum of Squared Error (SSE).
\end{itemize}

\subsection{External validity indices}

External validity measures are used to compare the resultant clustering solution with the true clustering of the data, if available.
These indices are included in  {\bf EXCLUVIS}, as these are very useful for comparing the performance of different clustering 
techniques when the true clustering is known. Suppose $T$ is the true clustering of a dataset and $C$ is a clustering result
given by some clustering algorithm. Let $a$, $b$, $c$ and $d$ respectively denote the number of pairs of points belonging to
the same cluster in both $T$ and $C$, the number of pairs belonging to the same cluster in $T$ but to different clusters in $C$,
the number of pairs belonging to different clusters in $T$ but to the same cluster in $C$, and the number of pairs belonging to
different clusters in both $T$ and $C$. Then the external validity indices implemented in (EXCLUVIS) are defined as follows.

\subsubsection{Minkowski index}

The Minkowski index \cite{ag03} $M(T,C)$ is defined as
\begin{equation}
M(T,C) = \sqrt{\frac{b + c} {a + b}}.
\end{equation}
Lower values of the Minkowski index indicate better matching between $T$ and $C$, where the minimum value $M(T,T)$ = 0.

\subsubsection{Adjusted Rand index}

The adjusted Rand index \cite{kw01} $ARI(T,C)$ is then defined as
\begin{equation}
ARI(T,C) = \frac{2 (a * d - b * c)} {(a + b) (b + d) + (a + c) (c + d)}.
\end{equation}
The value of $ARI(T,C)$ lies between 0 and 1; a higher value indicates that $C$ is more similar to $T$. Also, $ARI(T,T)$ = 1.

\subsubsection{Percentage of correctly classified pairs}

This index \cite{sua07} is computed as
\begin{equation}
P(T,C) = \frac{((a + d) \times 100)} {(a + b + c + d)}.
\end{equation}
The value of $P(T,C)$ lies between 0 and 100; a higher value indicates that $C$ is more similar to $T$. Also, $P(T,T)$ = 100.

\subsection{Internal validity indices}
The result of one clustering algorithm can be very different from another for the same input dataset as 
the other input parameters of an algorithm can substantially affect the behavior and execution of the algorithm. 
Internal validity indices are used to evaluate the quality of a clustering solution using the geometrical property of
the clusters, such as compactness, separation and connectedness. These indices may serve as an objective function
in order to determine the optimal clustering  structure of a dataset. To serve the purpose of validating a clustering 
solution, the following internal validity indices are implemented in {\bf EXCLUVIS}.

\subsubsection[J index]{$J$ index}

$J$ index \cite{bez81cmeans} is minimized by fuzzy C-means clustering. It is defined as follows:
\begin{equation}
J=\sum_{k = 1}^{K}\sum_{i = 1}^{n}u_{ki}^{m}D^{2}(Z_k , X_i),
\end{equation}
where $u_{ki}$ is the fuzzy membership matrix (partition matrix) and $m$ denotes the fuzzy exponent.
$D(Z_k , X_i)$ denotes the distance between the $k$th cluster center $Z_k$ and the $i$th data point $X_i$ . 
$J$ can be considered as the global fuzzy cluster variance. A lower value of $J$ index indicates more compact clusters. 
However, the $J$ value is not independent of the number of clusters $K$, i.e., if the value of $K$ increases, the $J$
value gradually decreases and it takes the minimum value 0 when $K = n$. It is possible to have a crisp version of $J$
when the partition  matrix $u$ has only binary values.

\subsubsection{Davies-Bouldin index}

Davies-Bouldin ($DB$) index \cite{dld79} is a function of the ratio of the sum of within-cluster scatter to between-cluster separation. The scatter within the $i$th cluster $S_i$ is computed as
\begin{equation}
S_i = \frac{1}{\mid{C_i}\mid}\sum_{\substack{x\in C_i}} D^2(Z_i , x).
\end{equation}
Here $|C_i|$ denotes the number of data points belonging to cluster $C_i$. The distance between two clusters $C_i$ and $C_j$, $d_{ij}$ is defined as the distance between the centers.
\begin{equation}
d_{ij}=D^2(Z_i , Z_j).
\end{equation}
The $DB$ index is then defined as
\begin{equation}
DB = \frac{1} {K}\sum_{\substack{i = 1}}^K R_i,
\end{equation}
where
\begin{equation}
R_i = \max_{j , j\neq i}\left\{{\frac{S_i + S_j}{d_{ij}}}\right\}.
\end{equation}
The value of $DB$ index is to be minimized in order to achieve proper clustering.

\subsubsection{Dunn index}

Suppose $\delta (C_i,C_j)$ denotes the distance between two clusters $C_i$ and $C_j$, and $\delta (C_i)$ denotes the diameter of cluster $C_i$; 
then any index of the following form falls under Dunn family of indices \cite{jcd74}:
\begin{equation}
DN = \min_{1\leq i \leq K}\left\{\min_{1\leq j \leq K ,j\neq i}\left\{\frac{\delta(C_i , C_j)}{\max_{1\leq k\leq K}\{\Delta(C_k) \}}\right\}\right\}.
\end{equation}
Originally Dunn used the following forms of $\delta$ and $\Delta$:
\begin{equation}
\delta(C_i,C_j) = \min_{x\in C_i , y\in C_j}\left\{D( x , y)\right\},
\end{equation}
and
\begin{equation}
\Delta (C_i) = \max_{x , y\in C_i} \left\{D(x , y)\right\}.
\end{equation}
Here $D(x,y)$ denotes the distance between the data points $x$ and $y$. A larger value of Dunn index implies compact and well-separated clusters. Hence the objective is to maximize Dunn index.

\subsubsection{Xie-Beni index}

Xie-Beni ($XB$) index \cite{xlg91} is defined as a function of the ratio of the total fuzzy cluster variance $ \sigma $ to the minimum separation $sep$ of the clusters. Here $ \sigma $ and $sep$ can be written as
\begin{equation}
\sigma = \sum_{k = 1}^K \sum_{i = 1}^n u_{ki}^2 D^2(Z_k , X_i),
\end{equation}
and
\begin{equation}
sep = \min_{k\neq l}\{ D^2(Z_k , Z_l)\}.
\end{equation}
$XB$ index is then written as
\begin{equation}
XB = \frac{\sigma}{n \times sep}.
\end{equation}
Lower value of $\sigma$ and higher value of $sep$ indicate that the partitioning is good and compact. Hence, the objective here is to minimize the $XB$ index for achieving proper clustering.

\subsubsection[I index]{${\cal I}$ index}

${\cal I}$ index \cite{us02} is defined as follows.
\begin{equation}
{\cal I} =\Bigg(\frac{1} {K} \times \frac{E_l} {E_k}\times D_k \Bigg)^p,
\end{equation}
where
\begin{equation}
E_k = \sum_{k = 1}^K \sum_{j = 1}^n u_{kj}D(Z_k , X_j),
\end{equation}
and
\begin{equation}
D_k = \max_{i , j = 1}^{K}\bigg \{D(Z_i , Z_j) \bigg \}.
\end{equation}
The different symbols used are as discussed earlier. $I$ index has three factors, namely $\frac{1}{K} , \frac{E_l} {E_k} , D_K$. 
The first factor tries to reduce value of index ${\cal I}$ as $K$ increases. The 
second factor consists of the ratio of $E_l$ to $E_K$; where $E_I$ is constant for a given dataset and value of $E_K$ decreased as value 
of $K$ increased. Hence, because of this term, index ${\cal I}$ increases as $E_K$ decreases. This, in turn, indicates that formation of more
clusters that are compact in nature would be encouraged. Finally, value of the third factor $D_K$~(which computes the maximum separation between
two clusters over all possible pairs of clusters), will increases with the
value of $K$. However, note that, maximum separation between two points in the dataset should be 
upper bound of this value. Thus, the three factors compete with and balance each
other. Contrast between the different cluster configurations is controlled by the power $p$.
It can be said that clustering is better if value returned by ${\cal I}$ index is high.

\subsubsection{Silhouette index}

Suppose $a_i$ represents the average distance of an assigned point $x_i$ from the other points of the cluster, and $b_i$ represents the minimum of the
average distances. Then the silhouette width $S_i$ of the point can be defined as follows.
\begin{equation}
S_i = \frac{b_i - a_i} {\max(a_i , b_i)}.
\end{equation}
Now, Silhouette index \cite{p87} ${\cal S}$ is the averaged silhouette width of all the data points.
\begin{equation}
{\cal S} = \frac{1} {n}\sum_{i = 1}^n S_i.
\end{equation}
Note that the value of the Silhouette index varies from -1 to 1, where a higher value indicates a better clustering result.

\section{System requirements}

The {\bf EXCLUVIS} package has been developed in {MATLAB 2009b}. Hence for running the software, a machine with {MATLAB R2009b} or higher is required. Note that {MATLAB} toolboxes
such as bioinformatics and statistical toolboxes are also essential in order to run {\bf EXCLUVIS}, as this package uses some features of those toolboxes. In terms of memory requirement, the machine must have at least 1 GB RAM. However 2 GB RAM is recommended for smooth performance. 

\section[Demonstration of the EXCLUVIS package]{Demonstration of the {\bf EXCLUVIS} package}

{MATLAB} GUI is used to develop this application package in order to give the users full flexibility to perform comparative studies of clustering algorithms and to visualize gene expression data without implementing the corresponding algorithms and the methods. In this application package, different graphical components such as Push Button, Radio Button, Edit Box, Static Text Box, Pop-Up menu, Toggle Button, Table, Axes, Panel, Button Group, Labels, and Menus have been used. 

\subsection{Initial window}

\begin{figure}[t!]
	\centering
	\includegraphics[width=\textwidth]{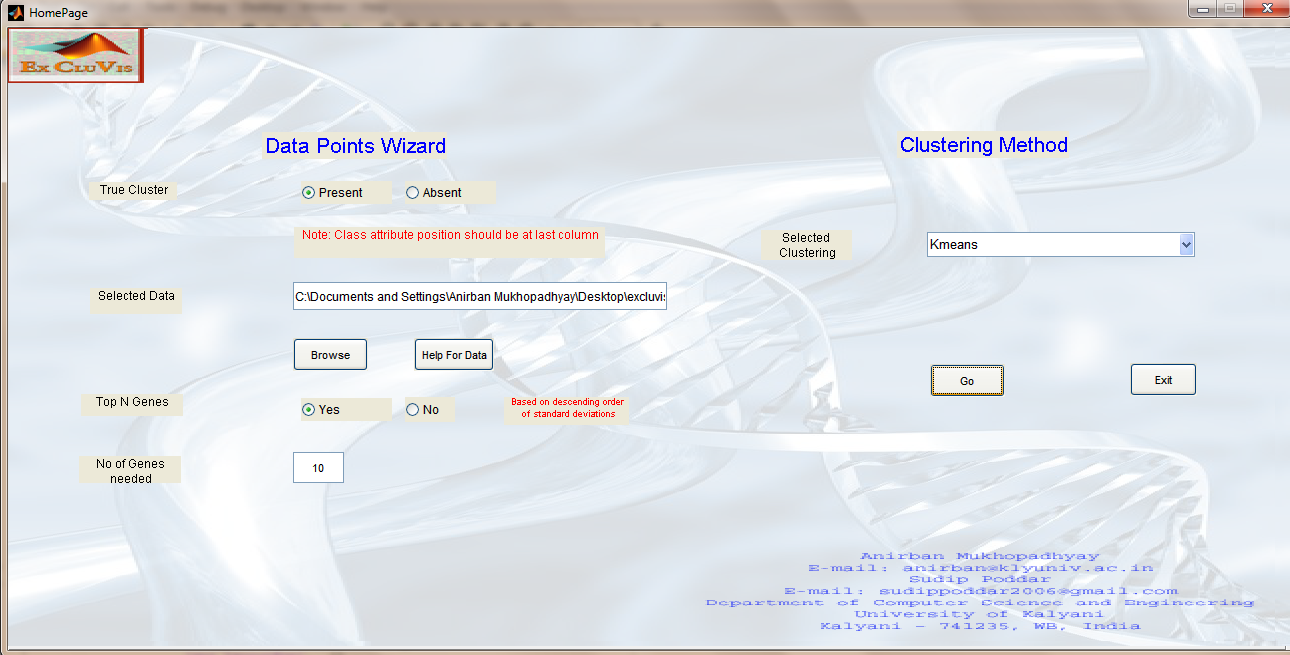}
	\caption{HomePage of {\bf EXCLUVIS}.}
	\label{Figure:1}
\end{figure}  
The initial window of this application package is shown in Figure~\ref{Figure:1}. For selecting the dataset, a browser button is 
included in the ``HomePage window''. This window can be invoked by typing {\bf HomePage} in the {MATLAB} prompt and 
pressing the return key. The dataset needs to be pre-processed in order to do analysis and the format of the dataset will be 
displayed by clicking on ``Help for Dataset'' button. The dataset should be real and no missing value is allowed. If true 
clustering exists (previous knowledge of the dataset) then the proper column number of class attribute needs to be specified in 
the edit box. The initial label vectors are saved in a directory for further analysis such as for finding the values of external 
validity indices. Top $N$ genes can also be selected from the input dataset by specifying the value of $N$.
For selecting the top genes, at first the variance of each gene is calculated and the genes are sorted in descending order of their variances. 
Thereafter, the top $N$ genes with highest variances are selected. The normalization is done in such a way that the minimum and 
maximum values of each row are mapped to default mean and standard deviation of 0 and 1, respectively. It is also assumed that 
the dataset has only finite real values, and that the elements of each row are not all equal. The selected dataset as well as the 
top genes are saved in different text files. Now, for doing analysis, any of the clustering algorithms' ($K$-means, Fuzzy $C$-
Means, Hierarchical clustering, MocSvm clustering) window can be opened by selecting the algorithm. If user wants to run all 
the clustering algorithms at the same time for doing the analysis, he/she can do it by choosing ``All clustering window'' option available at the initial HomePage window. 
For demonstration purposes, initially we have chosen the $K$-means window for doing the analysis on the selected dataset. ``Braintumor'' dataset
is selected for doing the analysis, which can also be found at \href{http://kucse.in/excluvis}{http://kucse.in/excluvis}. The 
dataset contains true clustering solution, and the column number of class attribute is 7130. We have selected top 100 genes 
for doing the analysis. In order to give the users a knowledge about the input information, tooltips are added in each field of
this application package and a tooltip is also shown in Figure~\ref{Figure:2} for demonstration purpose.

\subsection[K-means window]{$K$-means window}

\begin{figure}[t!]
	\centering
	\includegraphics[width=\textwidth]{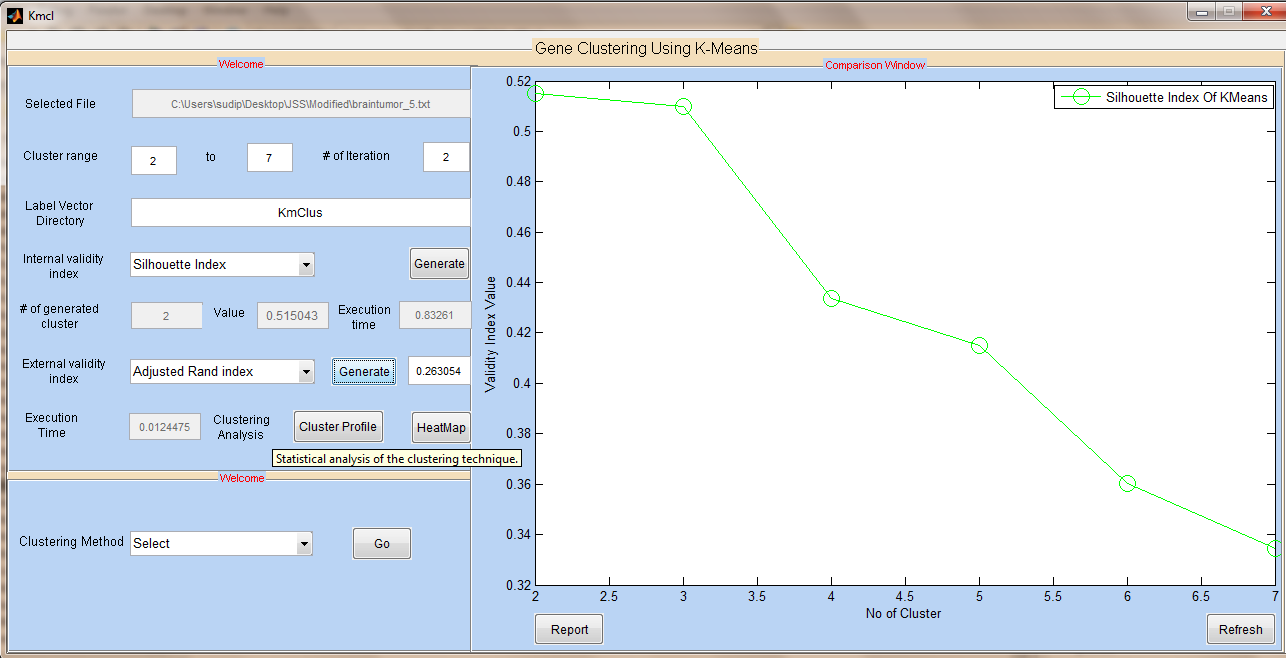}
	\caption{$K$-means clustering window.}
	\label{Figure:2}
\end{figure}

\begin{figure}[!h]
	\centering
	\includegraphics[width=\textwidth]{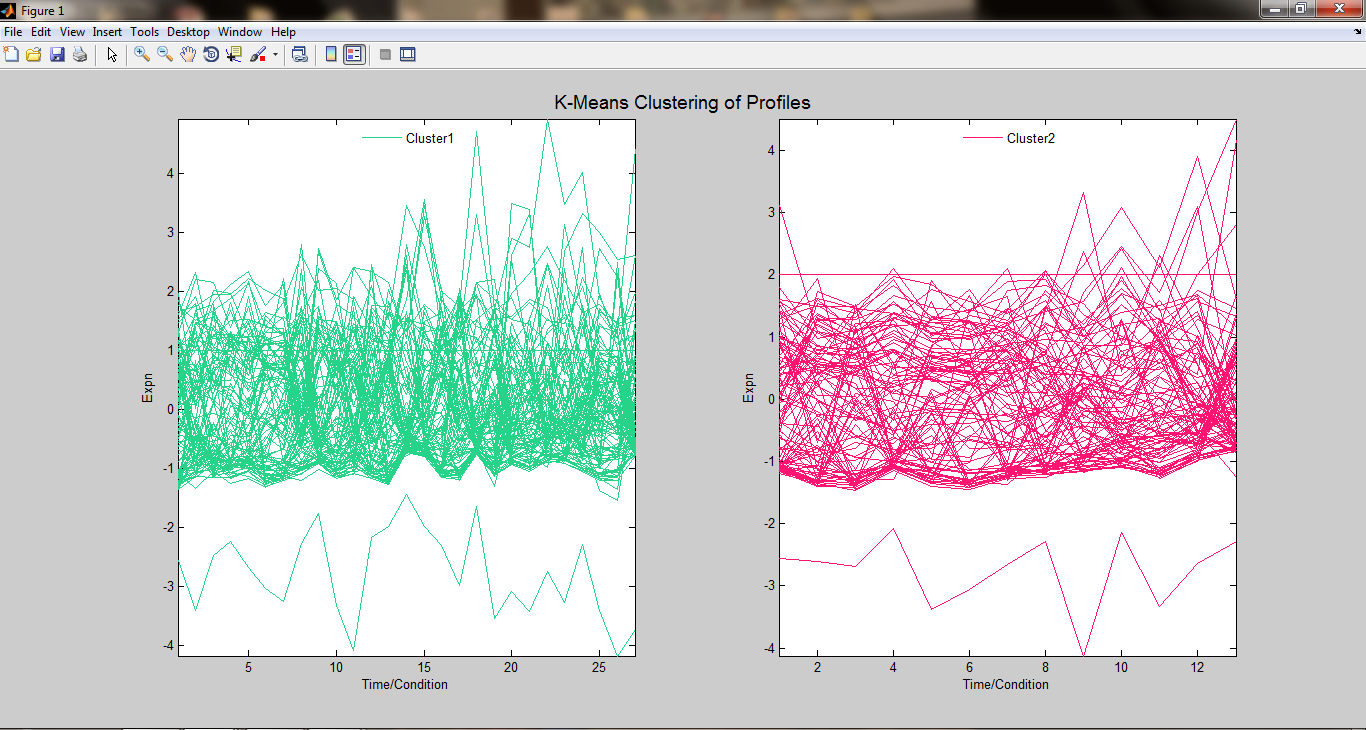}
	\caption{Heatmap of $K$-means clustering.}
	\label{Figure:3}
\end{figure}  

Initially, after opening the $K$-means window as shown in Figure~\ref{Figure:2}, name of the selected dataset in the previous
HomePage window automatically appears in the edit box of the $K$-means window. In order to cluster the selected dataset, user needs 
to give the values of probable range of clusters, and the number of iterations for determining the proper number of clusters. An internal validity index needs to be selected for finding the optimal number of clusters. Larger value indicates better result for Dunn Index, Silhouette Index, and {\cal I} index, whereas smaller value indicates better result for $J$ Index, Davies-Bouldin Index, and Xie-Beni Index. Default values are given in few fields, for example, 2 for number of iterations, label vector saved directory (here KmClus), etc. The generated number of clusters and the value of the chosen internal index will also appear in the edit box after clicking on the `Generate' button. In addition, a line graph is plotted on the plot field, where the $X$-axis denotes the range of number of clusters and the $Y$-axis denotes the generated validity index value for the corresponding number of clusters. For each number of cluster value, the corresponding index value is also marked with a unique marker on the graph. Legend on 
the plot indicates the plots of the internal validity indices used and the corresponding algorithm. Heatmap is also included to represent the level of expression of many genes across a number of comparable samples. Moreover, profile plot is added for showing the normalized gene expression values of the genes of each cluster with respect to the time points. The plot
window can be cleared by clicking on the `Refresh' button. For finding the time complexity, CPU running time is also displayed in the execution time field.
The procedure used for determining the value of the internal validity index (maximum or minimum depending upon internal validity index used) and number of cluster generated is as follows.

\begin{itemize}
	\item Suppose range of the given number of clusters is from 2 to 7 and the number of iterations is 2.
	\item Dunn index, Silhouette index, {\cal I} index, $J$ index, Davies-Bouldin index, and
	Xie-Beni Index are used as internal validity indices. Suppose Silhouette index is selected.
	\item  For iteration 1, index values are generated for each cluster value. Among all values, maximum~(minimum) one is chosen and corresponding
	label vector is also saved in a user specific directory. In our example, 6 different values are generated for Silhouette index, among them the maximum one is chosen and corresponding label vector is also saved in the specified directory.
	\item Similarly in the next iteration, another maximum value is selected and corresponding label vector is also saved. Finally, the maximum value is chosen among 2 values generated in 2 iterations 
	and corresponding label vector is retained in the assigned directory (KmClus) for further analysis.
\end{itemize}

\begin{figure}[t!]
	\centering
	\includegraphics[width=\textwidth]{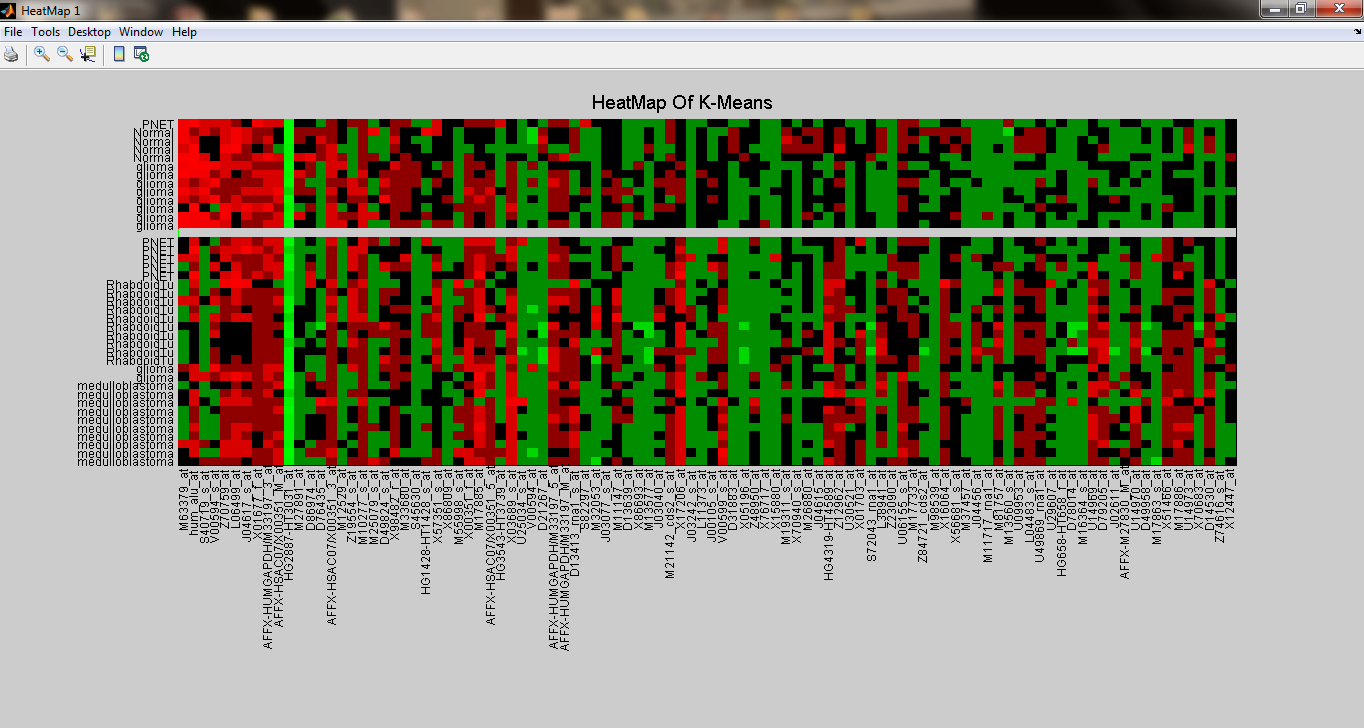}
	\caption{Profile plot of $K$-means clustering.}
	\label{Figure:4}
\end{figure}

Generated value of external validity indices, such as for Adjusted Rand index is also shown in Figure~\ref{Figure:2} along with the execution time. In order to compare the results of all clustering solution, a Report table window is also integrated in this package to store the number of clusters generated,
final index value as well as execution time for each clustering algorithm used.  The Report table can also be seen any time by clicking on the `Report' button. Moreover for clustering analysis, heatmap and 
profile plots are also incorporated in this application package. In heatmap, $X$-axis denotes time/conditions and $Y$-axis denotes the name of the genes and in each case, each cluster is separated by a separator in the heatmap. In Figure~\ref{Figure:3}, heatmap of $K$-Means clustering solution is shown, whereas profile plot of the clustering solution is shown in Figure~\ref{Figure:4}. This package is designed in such a way that in profile plot, profile of each cluster is drawn in a single window with different colors. Cluster value is also shown in each figure with the help of a legend.

\begin{figure}[!h]
	\centering
	\includegraphics[width=\textwidth]{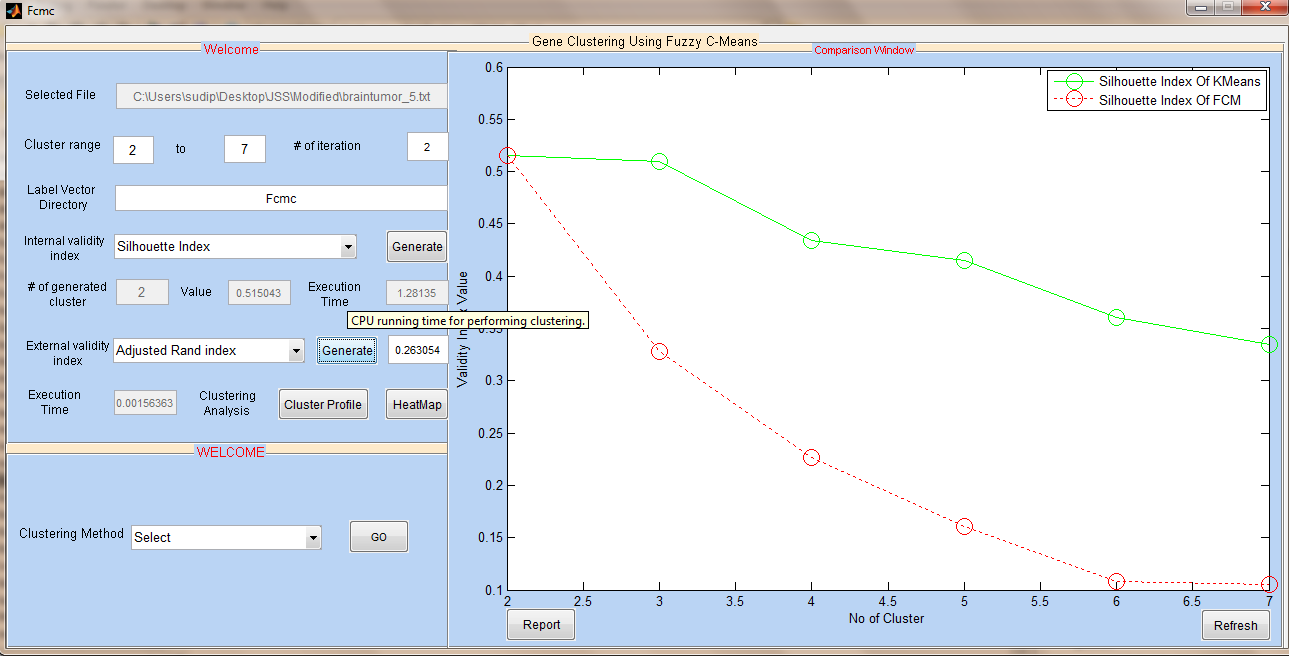}
	\caption{Fuzzy $C$-means clustering window.}
	\label{Figure:5}
\end{figure} 
\subsection[Fuzzy K-means window]{Fuzzy $C$-means window}

\begin{figure}[!h]
	\centering
	\includegraphics[width=\textwidth]{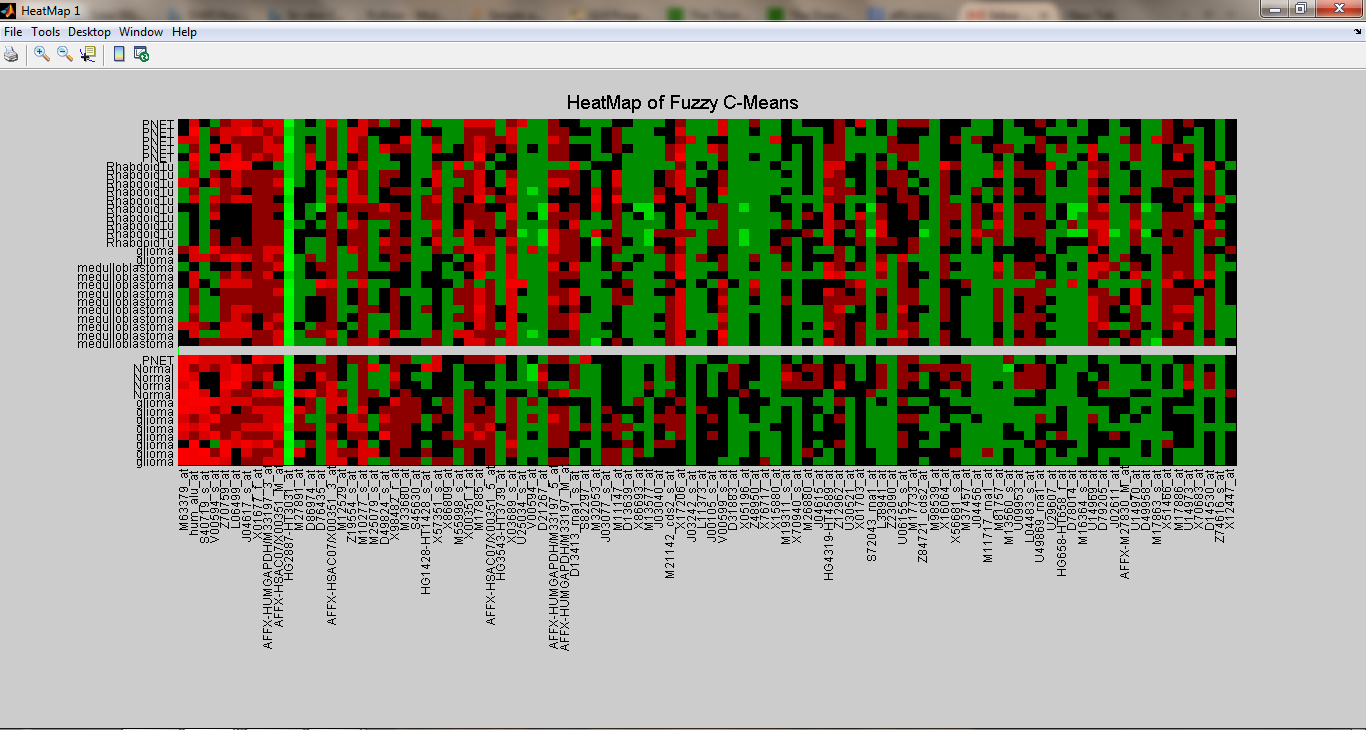}
	\caption{Heatmap of fuzzy $C$-means clustering.}
	\label{Figure:6}
\end{figure}

\begin{figure}[t!]
	\centering
	\includegraphics[width=\textwidth]{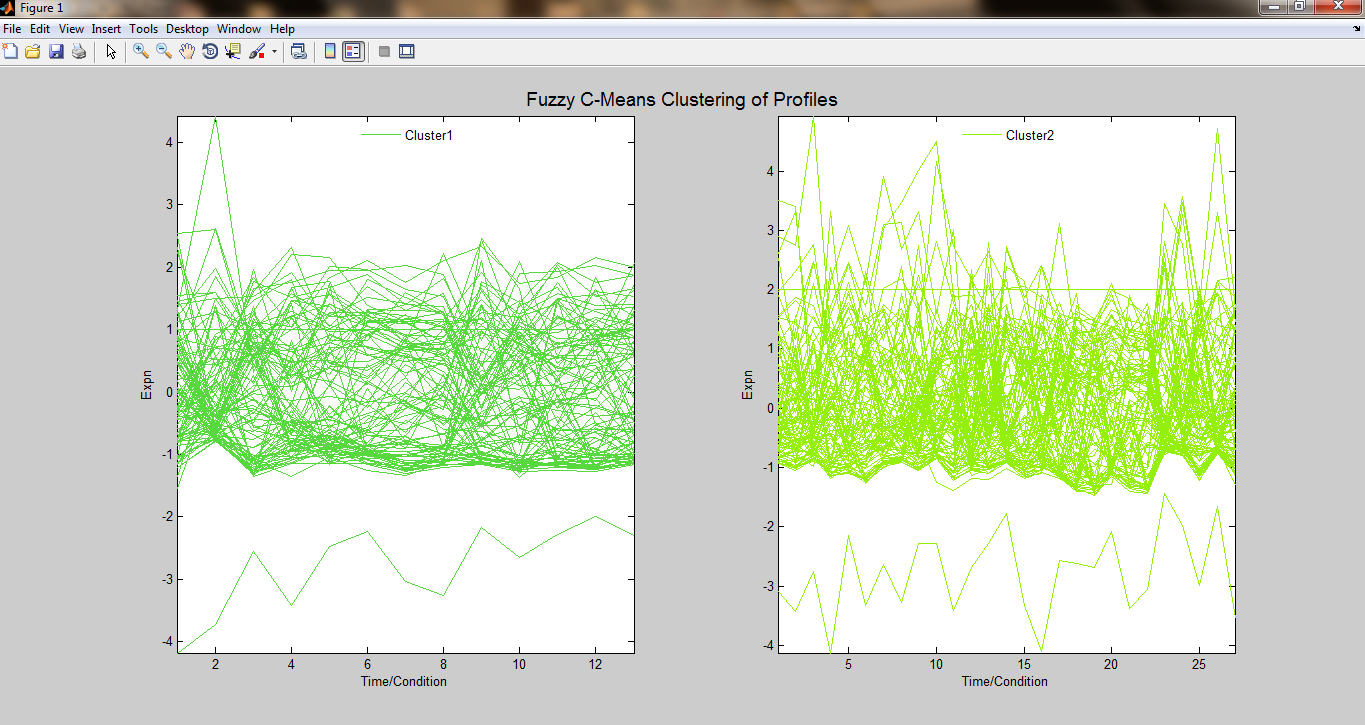}
	\caption{Profile plot of fuzzy $C$-means clustering.}
	\label{Figure:7}
\end{figure}

The {MATLAB} GUI code is written in such a way that all the values given by the user in the $K$-Means window are automatically populated in the corresponding fields of Fuzzy $C$-Means window, 
except the label vector saved directory (default values are given). Users also have the flexibility to change the value of any field in any time. Users can also find the values of internal as well as external 
validity indices by clicking on the appropriate buttons. All the generated values are
immediately saved in the Report table window. The line graphs having different colors along with the legends are also appended with
the previous plot in the plot window. Fuzzy C-Means window is shown in Figure~\ref{Figure:5}, whereas heatmap and profile plot windows are shown in Figure~\ref{Figure:6} and Figure~\ref{Figure:7}, respectively. Next for doing the analysis further, we select the Hierarchical clustering window.

\subsection{Hierarchical clustering}

\begin{figure}[t!]
	\centering
	\includegraphics[width=\textwidth]{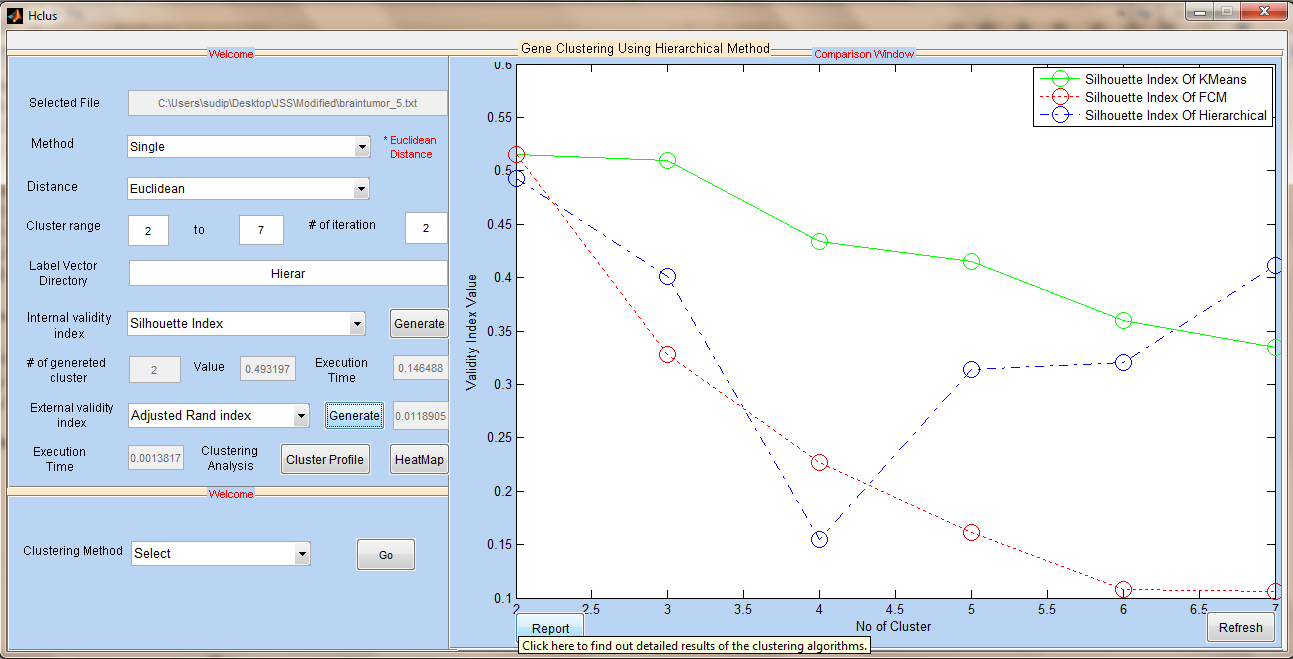}
	\caption{Hierarchical clustering window.}
	\label{Figure:8}
\end{figure}

\begin{figure}[t!]
	\centering
	\includegraphics[width=\textwidth]{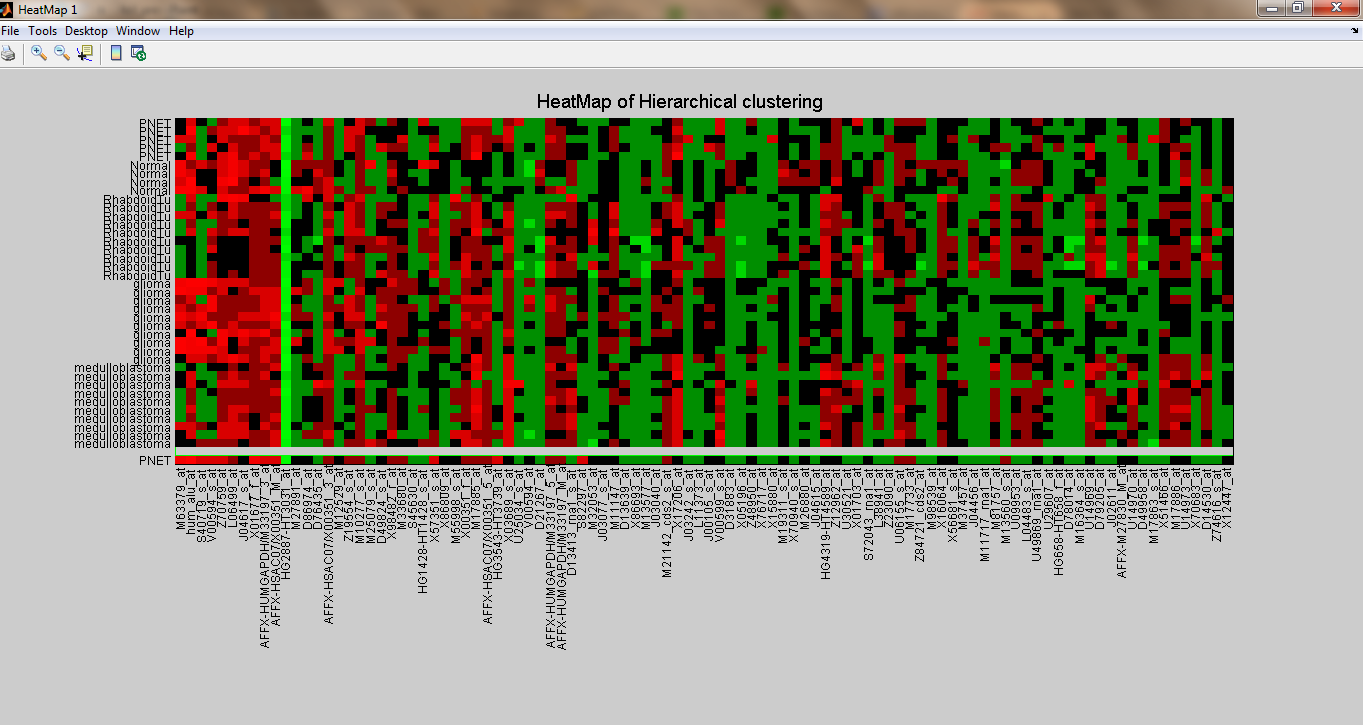}
	\caption{Heatmap of hierarchical clustering window.}
	\label{Figure:9}
\end{figure}

\begin{figure}[t!]
	\centering
	\includegraphics[width=\textwidth]{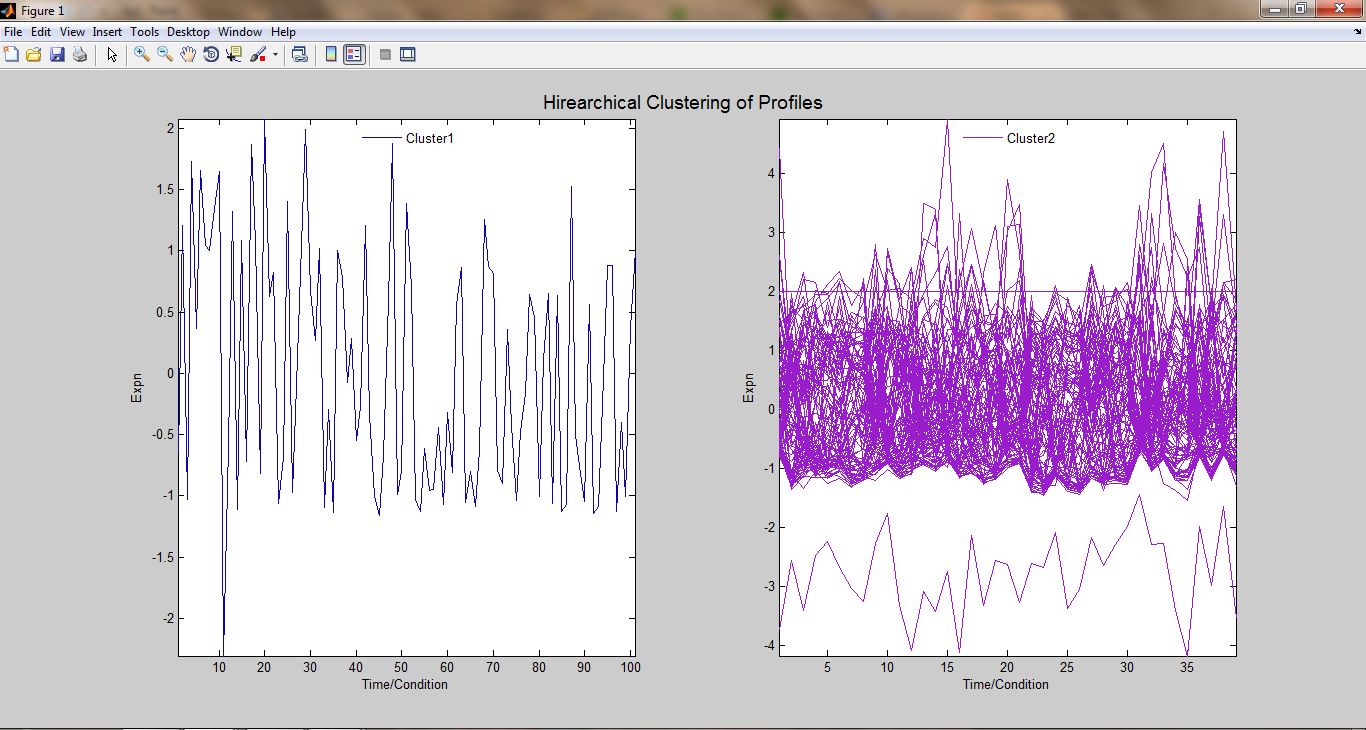}
	\caption{Profile plot of hierarchical clustering.}
	\label{Figure:10}
\end{figure}

After opening this window, selected data points, range of clusters, number of iterations, internal as well as external validity index field values, which are already set by the user in previous windows are automatically populated in the corresponding fields of this window as mentioned 
previously. To run this clustering algorithm, we added two extra fields  in this window, one field is `Distance' metric and another one is `Method' field. 
Distance metric is used for determining the choice of merging the 
cluster at each iteration. A distance metric need to be selected among Euclidean, Seuclidean, Cityblock, Mahalanobis, Minkowski, Cosine, Correlation, Spearman, Hamming, Jaccard and Chebychev. 
Note that, the distance function is chosen based on the linkage criteria. Among various linkage method available in this package, the user needs to select a method~(which is compatible with the distance metric)
for determining the distance between a cluster pair, when groups are formed. Available linkage methods are Single, Complete, Average, Weighted, Centroid, Median and Ward. Figure~\ref{Figure:8} shows the Hierarchical clustering window after calculating the values of internal and external validity indices. Note that, these values are also saved in Report table window and line graphs are drawn in the plot window with different colors and markers. Selected internal validity index as well as used algorithm are also labelled in the plot window using legend. Heatmap and profile plots of this clustering are shown in Figure~\ref{Figure:9} and Figure~\ref{Figure:10}, respectively.

\subsection{MocSvm clustering}

In this clustering technique, the genetic algorithm parameters which are additionally included are population size, probability of 
crossover (Pcrossover), probability of mutation (Pmutation), membership threshold of the point 
($\alpha$), threshold of fuzzy majority voting ($\beta$) and Weight. Generally population size is chosen by the user, whereas crossover is the exchange of genetic information that takes place between randomly selected parent chromosomes. Mutation is the random alteration in the genetic structure for introducing genetic diversity into the population of solutions. Both are probabilistic operations and generally the value of crossover probability is kept high and mutation probability is kept low.
\begin{figure}[t!]
	\centering
	\includegraphics[width=\textwidth]{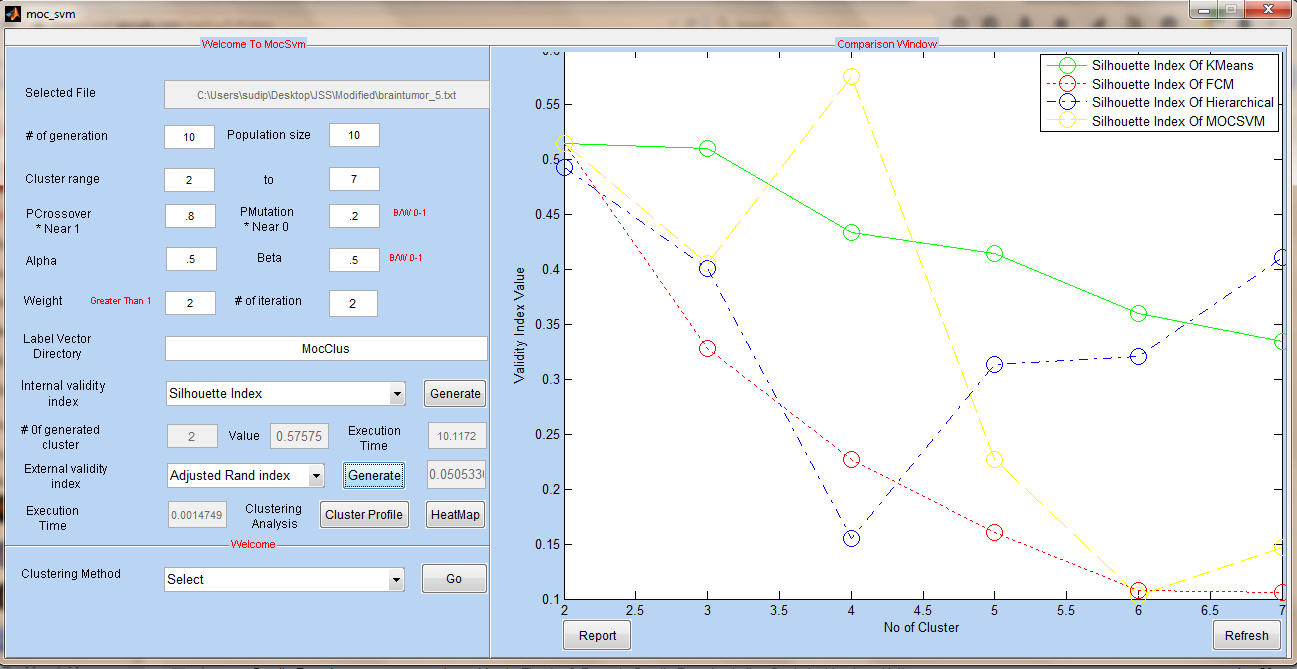}
	\caption{MocSvm clustering window.}
	\label{Figure:11}
\end{figure}
\begin{figure}[!h]
	\centering
	\includegraphics[width=\textwidth]{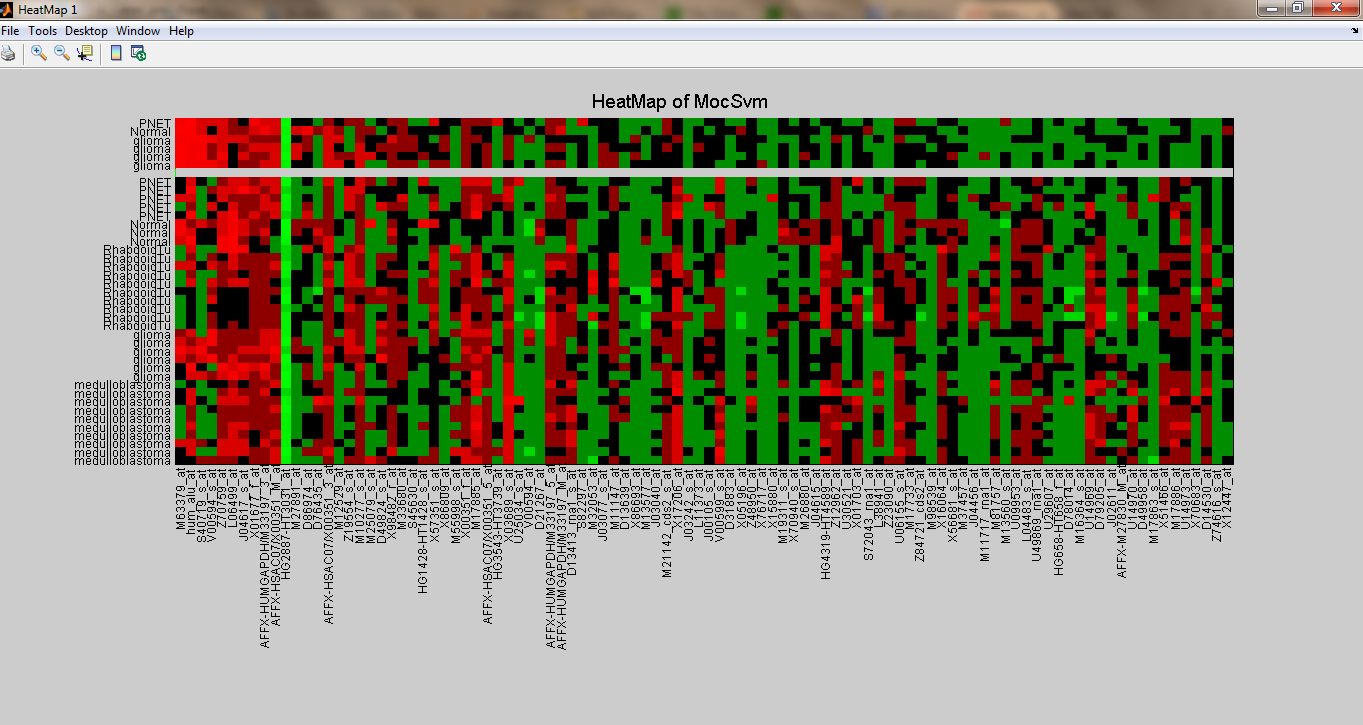}
	\caption{Heatmap of MocSvm clustering.}
	\label{Figure:12}
\end{figure}
In this clustering, the size of the training set depends on $\alpha$ and $\beta$. The size of the training set decreases by increasing the value of $\alpha$ and $\beta$. On the other hand, the size of the training set increases by decreasing the value of $\alpha$ and $\beta$. Generally the values of both parameters are set to 0.5 for finding a good solution. Generated internal as well as external validity indices values are saved in the
\begin{figure}[!ht]
	\centering
	\includegraphics[width=\textwidth]{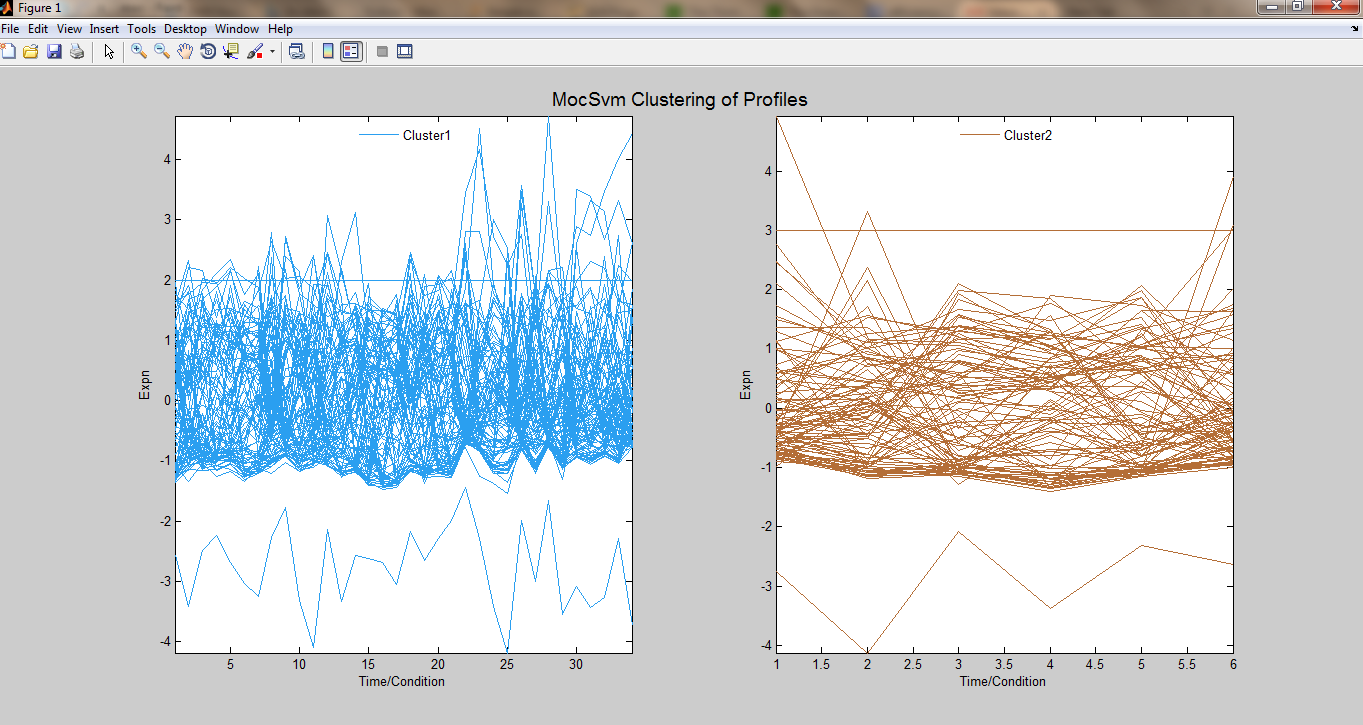}
	\caption{Profile plot of MocSvm clustering.}
	\label{Figure:13}
\end{figure}
report table window as mentioned previously. For further analysis, generated label vectors are also saved in the specified 
directory. Note that, for comparison purpose, line graph is also appended in the plot window with different marker as well as
different color. This graphical window is shown in Figure~\ref{Figure:11}. The heatmap and profile plot of this clustering solution are also demonstrated in Figure~\ref{Figure:12} and Figure~\ref{Figure:13}, respectively.

\subsection{Report Table Window}
\begin{figure}[t!]
	\centering
	\includegraphics[width=\textwidth]{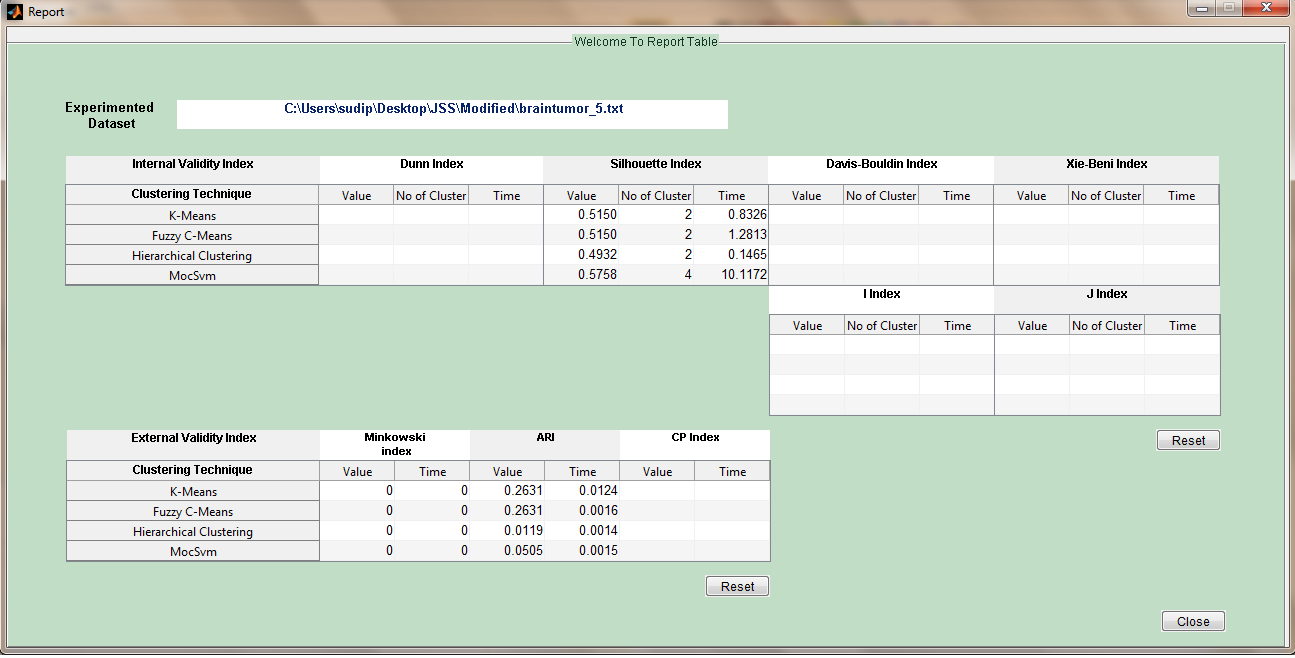}
	\caption{Report table window.}
	\label{Figure:14}
\end{figure}

This window contains two tables, one table is used for storing the generated  internal validity index values of an algorithm and another one is used for storing the generated values of external validity indices. Each internal validity index column is divided into three parts, namely i) generated values~(maximum or minimum depending on the used validity index), ii) number of generated clusters, and iii) time taken by cpu for running the algorithm. On the other hand, external validity index column is divided into two sub-columns, one is for the generated value and another is for storing CPU running time. If the dataset does not contain any true clustering solution, then external table does not appears in the Report table. During simulation, the values generated in each clustering window are automatically populated in the corresponding field of this window. Figure~\ref{Figure:14} shows an example of Report table populated with the corresponding generate values. Users can also close the Report table window any time by clicking on the `close' button.

\subsection{All clustering running window}

\begin{figure}[!h]
	\centering
	\includegraphics[width=\textwidth]{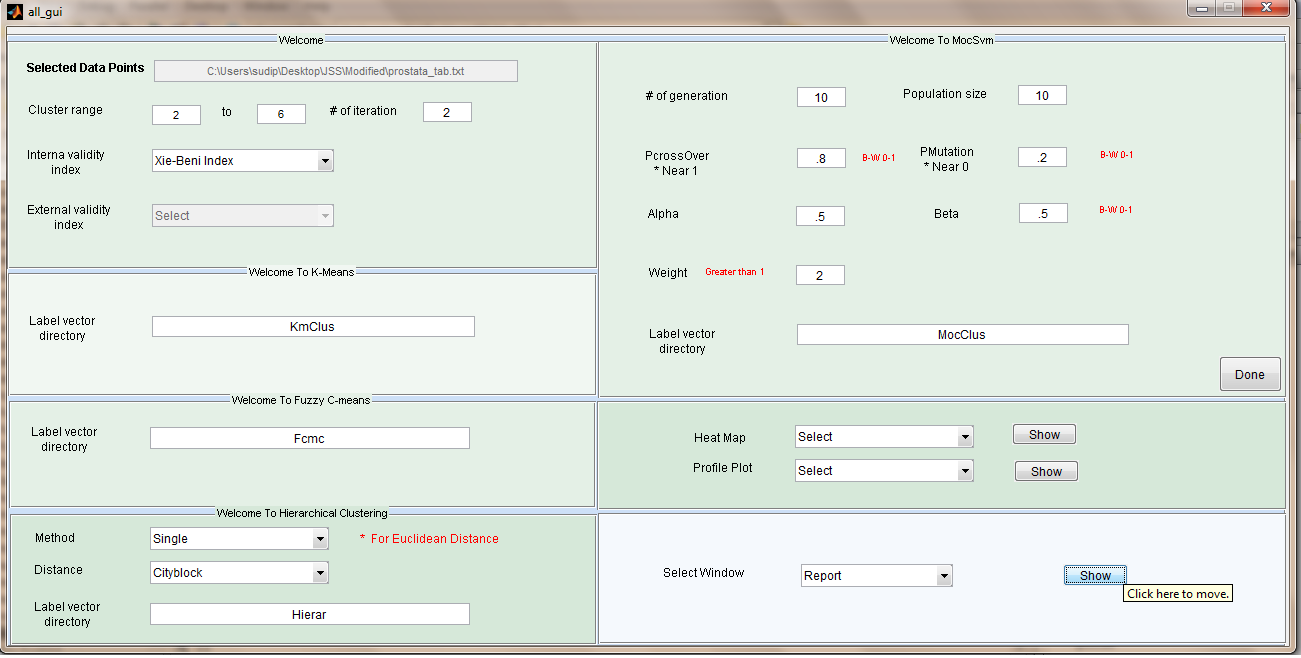}
	\caption{All cluster running window.}
	\label{Figure:15}
\end{figure}

Till now, we have demonstrated the clustering windows one by one and explained all the features of this application package 
by running them separately. However, in HomePage, there is also an option of choosing `All clustering window'. This window enables an user to run all the clustering algorithms at the same time, without running each clustering algorithm individually. This graphical window is shown in Figure~\ref{Figure:15}.
In this window, the user needs to give values of possible range of clusters, number of iterations, desired internal validity indices and external validity indices (if true clustering exists) only once for running all the clustering algorithms. For taking inputs, these fields
are grouped in a common panel of this window. If true clustering information does not exist, then external validity field button becomes inactive. $K$-Means, Fuzzy $C$-Means, Hierarchical Clustering, 
and MocSvm have different panels in this window for taking algorithm specific inputs. Default values are given to some fields of 
\begin{figure}[t!]
	\centering
	\includegraphics[width=\textwidth]{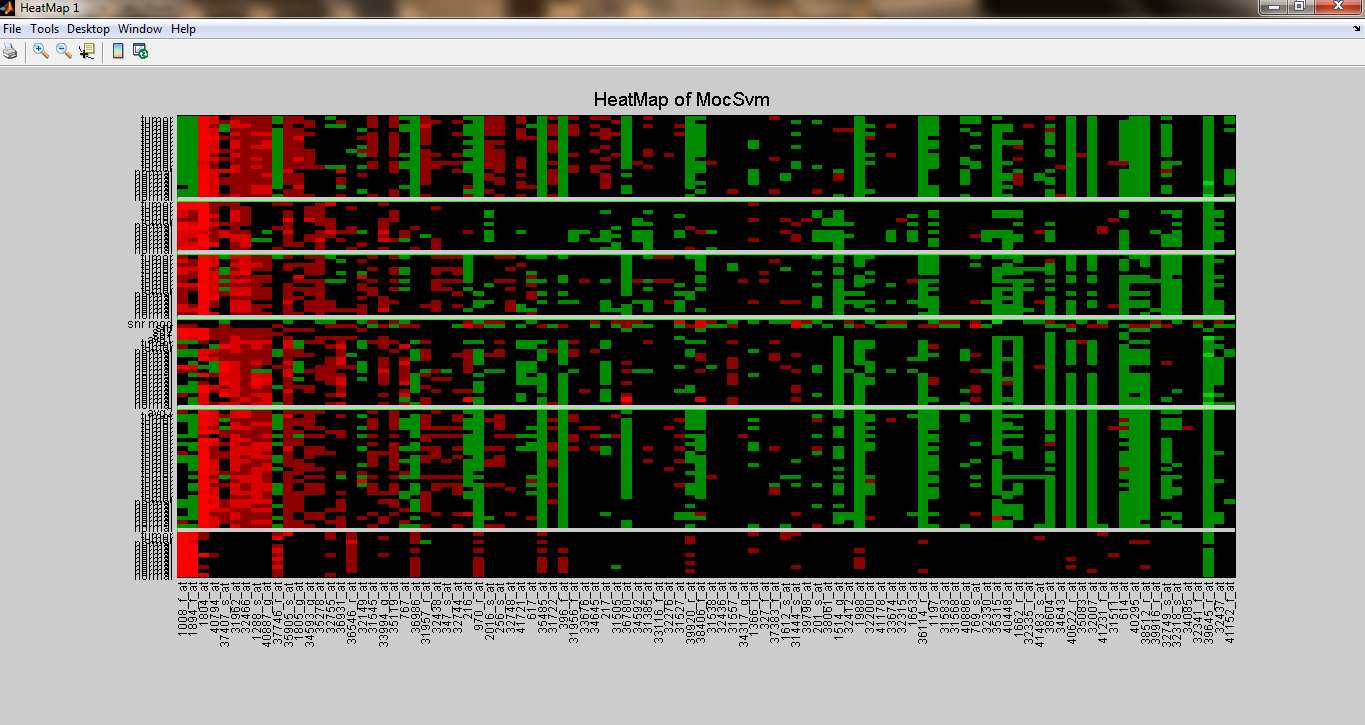}
	\caption{Heatmap of MocSvm clustering from ``all clustering running window''.}
	\label{Figure:16}
\end{figure}
\begin{figure}[h!]
	\centering
	\includegraphics[width=\textwidth]{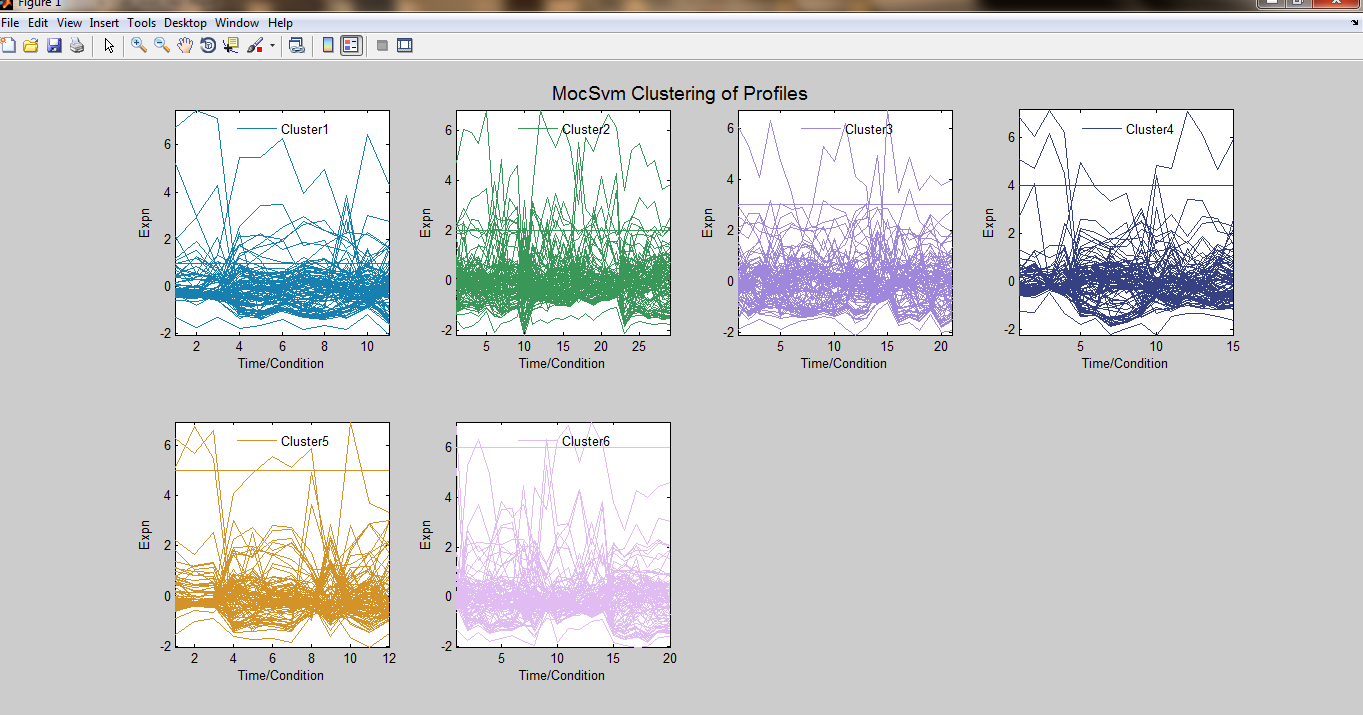}
	\caption{Profile plot of MocSvm clustering from ``all clustering running window''.}
	\label{Figure:17}
\end{figure}
this window, like other clustering windows. Input information of each field can be found by placing mouse on top of that field. The background code is written in such a way that after clicking on `Done' button, it first opens the $K$-Means window in invisible mode. Then all the input values given by the user in ``All clustering window", are automatically filled up in all the corresponding fields of $K$-Means window. Next it automatically creates the values of internal as well as external indices, and saves those generated values in the Report table. \begin{figure}[!ht]
	\centering
	\includegraphics[width=\textwidth]{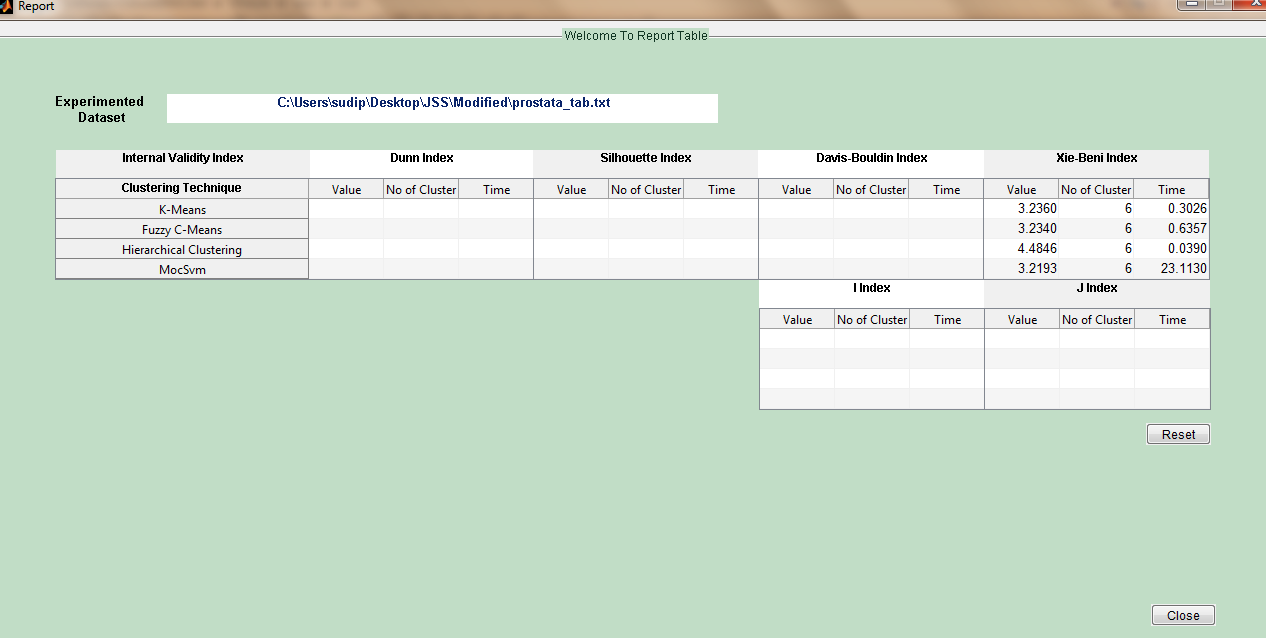}
	\caption{Report table window for running all clustering algorithms.}
	\label{Figure:18}
\end{figure}
Similarly other clustering algorithms are executed and the generated index values are stored in the Report table. Here for demonstration purpose we used the prostate (Singh {\em et al}.) dataset. In this window, users also has the option to visualize the solution of each clustering algorithm using heatmap or profile plot. Since prostate dataset does not contain any true clustering information, external validity index field remains inactive in all the clustering windows. The detailed results of all clustering algorithms are also shown in Figure~\ref{Figure:18} as Report table. 
For demonstration purpose, heatmap and profile plot of MocSvm clustering are shown in Figure~\ref{Figure:16} and Figure~\ref{Figure:17}, respectively.

\subsection{Comparison of clustering algorithms}
{\bf EXCLUVIS} package is also suitable for comparison of different clustering algorithms for a particular dataset. Comparison can be done both visually as well as numerically. For visual comparison, the user can produce the validity index plots with respect to different number of clusters for different algorithms, and then these plots can be visualized for comparison as shown in Figure~\ref{Figure:11}. It is evident from the figure that, different clustering algorithms provide best (maximum) value of Silhouette index for different number of clusters. $K$-means, Fuzzy $C$-means, hierarchical clustering and MocSvm provide the maximum values of Silhouette index when the numbers of clusters are 2, 2, 2 and 4, respectively. It is clear that MocSvm provides the maximum value of Silhouette index. Moreover, the different algorithms can be compared based on the best values obtained for different cluster validity indices using the report table window as shown in Figure~\ref{Figure:14}. This window shows the values of the validity indices for different algorithms. This helps direct comparison of different algorithms. It is evident from the figure that MocSvm provides the maximum value of Silhouette index (0.5758). However, MocSvm takes the maximum time (10.1172 seconds) among all the algorithms. Furthermore, to facilitate one-click comparison of different algorithms, {\bf EXCLUVIS} also has an ``all clustering running window'' (Figure~\ref{Figure:15}) which takes input for each clustering algorithm and then runs all the algorithms on one click and produces the results as desired (see Figure~\ref{Figure:18}). This way, {\bf EXCLUVIS} not only helps running individual clustering algorithms, but also facilitates comparison of different algorithms.



\section{Conclusion}

In this article, we have presented an effective and user-friendly application tool for Gene clustering. We have developed this analytical tool and software using {MATLAB} toolboxes in such a way that users can also include new algorithms in it, if required. It is also possible to include new validity indices~(internal or external) in this application package.
{\bf EXCLUVIS} has several features that make it a potentially useful tool for a community of researchers and developers. Researcher can also visualize the results using the features like heatmap and cluster profile plot available in {\bf EXCLUVIS}.
In future, we have plan to upgrade the software package in multiple ways. For example, we have plans to include more clustering algorithms in this application package. Moreover, some other newly developed validity indices may be incorporated. As the software is focused towards analyzing gene expression data, we have also plan to include biological validation options with the help of gene ontology.

\scriptsize 
\bibliographystyle{./IEEEtran}

\end{document}